\documentclass[twocolumn,prb,aps,amsmath,amssymb,floatfix]{revtex4-1}
\pdfoutput=1

\usepackage{graphicx}
\usepackage{epstopdf}
\usepackage{epsfig}
\usepackage{amsbsy}
\usepackage{color}

\begin{document}
\title{Confined states in quantum dots defined within finite flakes of bilayer graphene: coupling to the edge, ionization threshold
and valley degeneracy}

\author{D. P. \.Zebrowski, E. Wach, and B. Szafran}
 \address{AGH University of Science and Technology, \\ Faculty of Physics and Applied Computer Science,
al. Mickiewicza 30, 30-059 Krak\'ow, Poland}

\date{\today}

\begin{abstract}
We study quantum dots defined by external potentials within finite flakes of bilayer graphene
using the tight-binding approach.
We find that in the limit of large flakes containing zigzag edges the dot-localized energy levels
appear within the energy continuum formed by extended states. In consequence no ionization threshold for the carriers
contained within the dot exists. For smaller flakes with zigzag boundaries the dot-localized energy levels
appear interlaced with the energy levels outside the flake, so in a charging experiment the electrons
will be added alternately to the dot area and to its neighborhood.
We demonstrate that for flakes with armchair boundaries only, an energy window accessible uniquely to the dot-localized
states is formed. Then a number of electrons can be added to the dot before the external states start to be occupied.
We also discuss coupling of the dot-localized states to the edge-states in the context of the valley
degeneracy lifting. Moreover, we extract smooth envelope wave functions from the tight binding solution and discuss their spatial symmetries. The coupling of the dot localized energy levels with reconstructed zigzag edges and atomic vacancies present within the layers is also considered.
\end{abstract}

\maketitle

\section{Introduction}

Graphene \cite{grar} quantum dots \cite{gqd,cons,neg} are generally considered attractive candidates for quantum information storage
and processing using spins of confined electrons \cite{cons,neg} because of their negligible hyperfine interaction with the carbon
lattice.\cite{neg}
Due to the gapless and linear form of the graphene band structure near the $K$ points of the Brillouin zone,\cite{walas}
electrons behave like relativistic Dirac particles, and as such evade confinement
by purely electrostatic means with transfer probability equal to 1 for the normal incidence
of the electron to the potential step.\cite{klein}
The spatial confinement is naturally present in small graphene flakes
which thus can serve as quantum dots.\cite{gqd}
However, for  small flakes of graphene the properties of the electron states
near the Fermi energy are largely determined by the form of the boundary,\cite{peeters,jaskolski,potasz,wurmetal} which is rather hard to be precisely
controlled. The continuum model predicts \cite{uwaga} formation of localized states within the energy continuum
in infinite layers of graphene for rotationally invariant potentials and non-zero angular momenta, for which
the electron current is never at a normal incidence to the edge of the dot area.

Electrostatic confinement that should allow for
fabrication of controllable devices with reproducible properties
can be formed in bilayer graphene \cite{review}
with external electric fields that  introduce inversion asymmetry between
the layers and open an energy gap in the band structure.\cite{gap}
In particular, the existence of stable-localized-states in bilayer graphene
quantum dots \cite{pereira,pereirar} was demonstrated.

The idea for electrostatic confinement in bilayer graphene  was recently exploited in experimental setups,\cite{dro,goo,yac}
in which charging of the quantum dot was studied. For small flakes ($85$ nm $\times$ 50 nm
as in Ref. \onlinecite{dro} for instance) the coupling of the confined states to the edges or charging of the edge
instead of the intentionally defined quantum dot can be expected.
The present paper investigates this issue.

We consider a finite flake of Bernal stacked bilayer graphene with a quantum dot defined in its inside
and edges of a zigzag and/or armchair form.
 Since the study requires a careful description of the boundaries we use the tight-binding Hamiltonian \cite{review} instead of its continuous low-energy approximation.\cite{pereira} We account for the dots diameter up to 50 nm and the
flakes of areas up to $80\times80$ nm$^2$.
We find that when the flake contains zigzag edges the localized states within the quantum dots are not separated by any energy gap from the external ones
and evidence of coupling between these states is found in the energy spectra.
For the flake containing the armchair edges only the
bound states are separated from the delocalized ones by a distinct energy gap. In this case
a finite ionization threshold exists, i.e. a finite energy that is required to remove the electron from the dot, like for quantum dots
in bulk 3D semiconductors. We find that the coupling to the edges lifts the valley degeneracy of the dot-localized states.
We also extract the smooth envelope functions from the rapidly varying tight-binding solutions near the Fermi energy,
and compare the symmetries of the calculated wave functions with the ones expected for the Dirac Hamiltonian
of the continuum approximation for circular potentials and infinite graphene plane.\cite{pereira}

\begin{figure}[ht!]
\hbox{
\hspace*{0.2cm}
\includegraphics[scale=0.28]{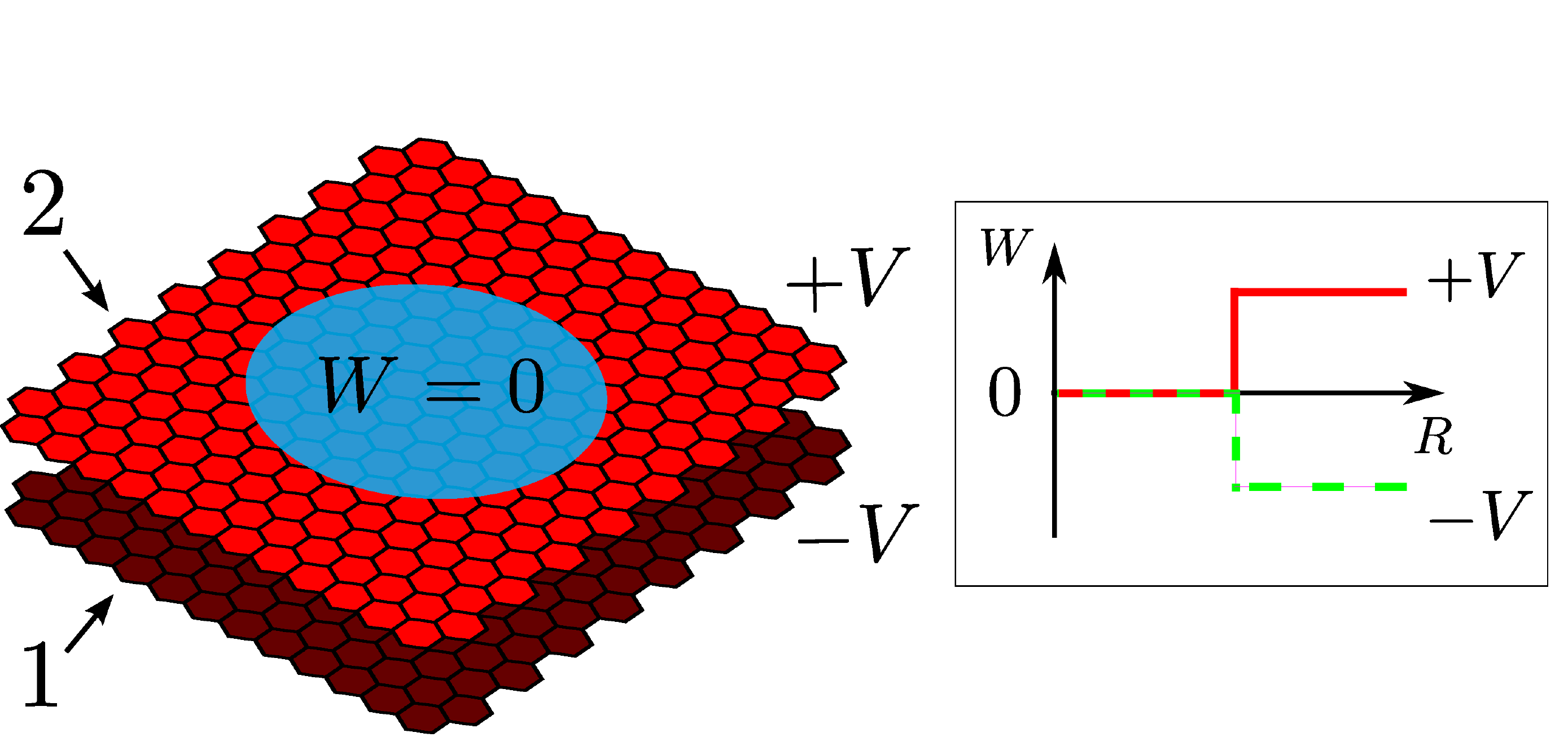}
}
\caption{Schematics of the considered bilayer system. An opposite
potential is applied to the two layers outside the circular central region
which forms a quantum dot
of radius $R$.  The plot at the right shows the
potential profile in function of the distance from the center of the dot for the upper (red curve) and lower (green curve) layers.
}
\label{schemat1}
\end{figure}
\begin{figure}[ht!]
\hbox{
\hspace*{0.6cm}
\includegraphics[scale=0.28]{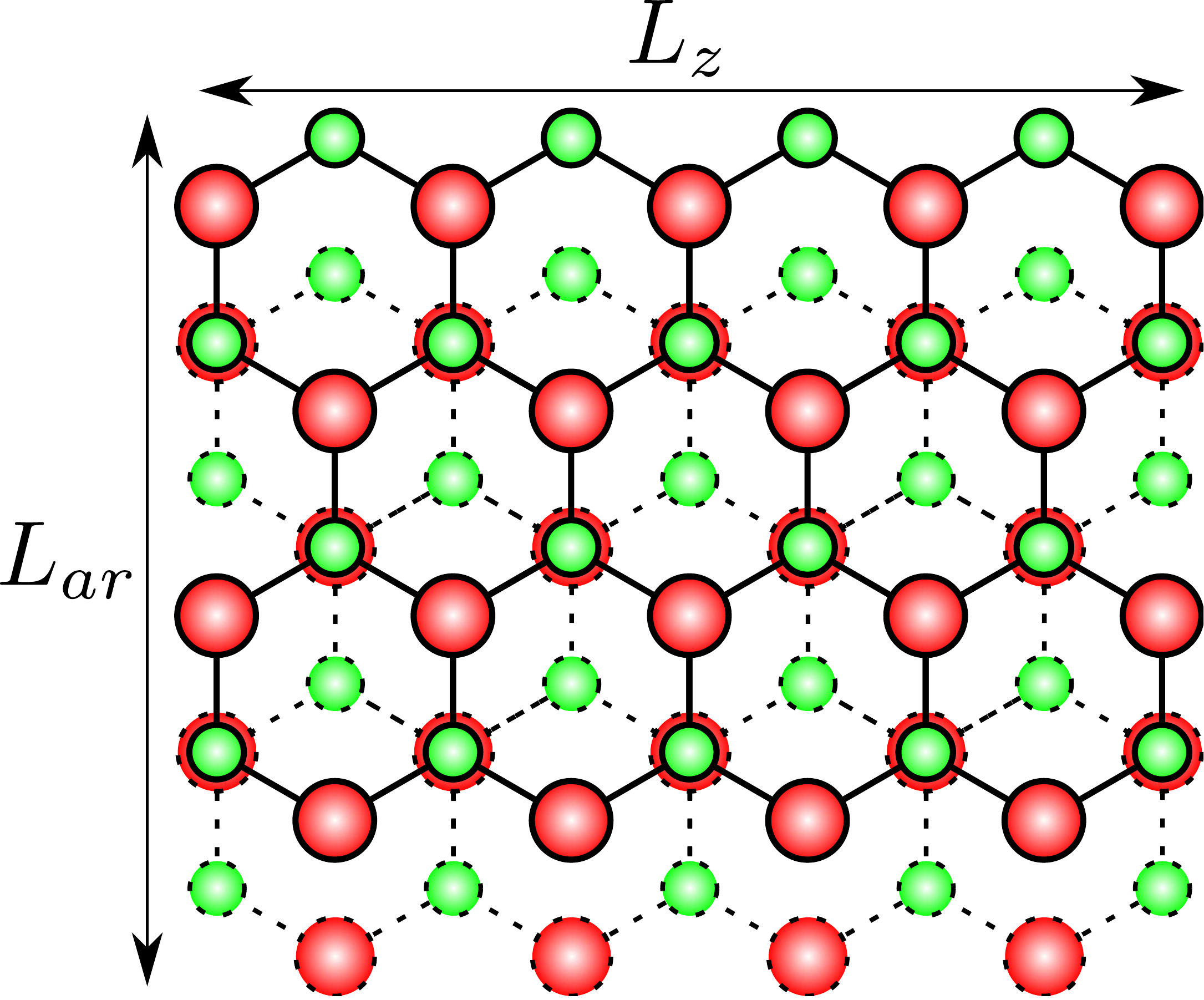}
          }
\caption{The carbon lattice for the rectangular flake of bilayer graphene in Bernal stacking (solid lines correspond to the upper layer, dashed lines to the lower). The  layers are composed of two sublattices $A$ (red), and $B$ (green). Subblatice $B$ of the upper layer ($2B$) couples to subblatice $A$ of the lower layer ($1A$).
 $L_z=\sqrt{3} N_z a$, and $L_{ar}=\frac{3}{2}a N_{ar}$ are the length of the zigzag and armchair edges,
with the lattice constant $a=0.142$ nm.
The  total number of atoms within the flake is $2(2N_z+1)N_{ar}$, where $N_z$ and $N_{ar}$ are the numbers
of atoms at the zigzag and armchair edges.}
\label{schemat2}
\end{figure}

\section{Theory}

We consider the tight-binding Hamiltonian \cite{review}
\begin{equation}
H=\sum_{i\neq j} t_{ij} (c_i^\dagger c_j+h.c.)+\sum_i W_i c_i^\dagger c_i,\label{eqh}
\end{equation}
where $c_i^\dagger$ $(c_i)$ is the particle creation (annihilation) operator at ion $i$,
and $t_{ij}$ is the nearest neighbor hopping parameter, with  $t_{ij}=-2.7$ eV for neighbors
within the same layer (see Fig. \ref{schemat2}) and $t_{ij}=-0.3$ eV for stacked pairs of ions of the two layers.\cite{review}
The potential at the atom $i$ is considered of a step-like circular form \begin{equation}W_i=W({\bf r}_i)=\left\{\begin{array}{cc} 0 & \mathrm{ for\; } |{\bf r}_i|<R \\
V & \mathrm{upper\; layer\; for \;} |{\bf r}_i|\ge R \\
-V & \mathrm{lower\; layer\; for\; } |{\bf r}_i|\ge R\end{array} \right.,\end{equation}
where $R$ is the radius of the dot (see Fig. \ref{schemat1}).
We have chosen for presentation the step-like confinement as  most convenient for discussion
of the localized versus delocalized states. Nevertheless, we have checked that the properties discussed below remain qualitatively the same for smooth
potentials.
We do not consider the electron spin in this work, hence when in the discussion we state that an eigenstate of Eq. (\ref{eqh}) is a non-degenerate (two-fold degenerate),
it implies a two-fold (four-fold) degeneracy including the spin degree of freedom.
The tight-binding approach applied here has been previously used for analysis of the spectra of graphene quantum dots including the statistics of energy levels far
from the charge neutrality point,\cite{els} which are outside the applicability of the low-energy Dirac Hamiltonian.

Formation of electrostatic confinement in bilayer graphene can be induced on either
electrons or holes also when the spatially modulated external potential is applied
to one of the layers only with the other kept at a constant potential. \cite{pereira}
In this work we assume that opposite potentials are applied to the two layers
which allows for conservation of the electron-hole symmetry for carefully chosen shape of the flake.
In experiments with bilayer graphene the potential within the two layers is routinely applied
in an independent manner,\cite{dbgb}
to control both the width of the electric-field induced energy gap and its position
with respect to the Fermi energy.
Fabrication
of split bottom gates for graphene nanostructures has recently been demonstrated in Ref. \onlinecite{sbg}.

Figure \ref{schemat2} shows the atomic structure of a rectangular flake within which the dot is
defined by external voltages. We assume the Bernal stacking and that the left and right
boundaries of the flake are armchair edges aligned vertically. The edges at the top and bottom
ends of the flake are zigzag edges with dangling vertical bonds.
For the assumed form of the flake with equal areas of the layers and
the adopted relative shift between them  the electron-hole asymmetry
which usually appears in bilayer graphene \cite{mucha} is absent and the
energy spectra
are perfectly symmetric with respect to the Fermi energy level $(E=0)$.

\begin{figure*}[ht!]
\hspace*{-1.2cm}
\includegraphics[scale=0.9 ]{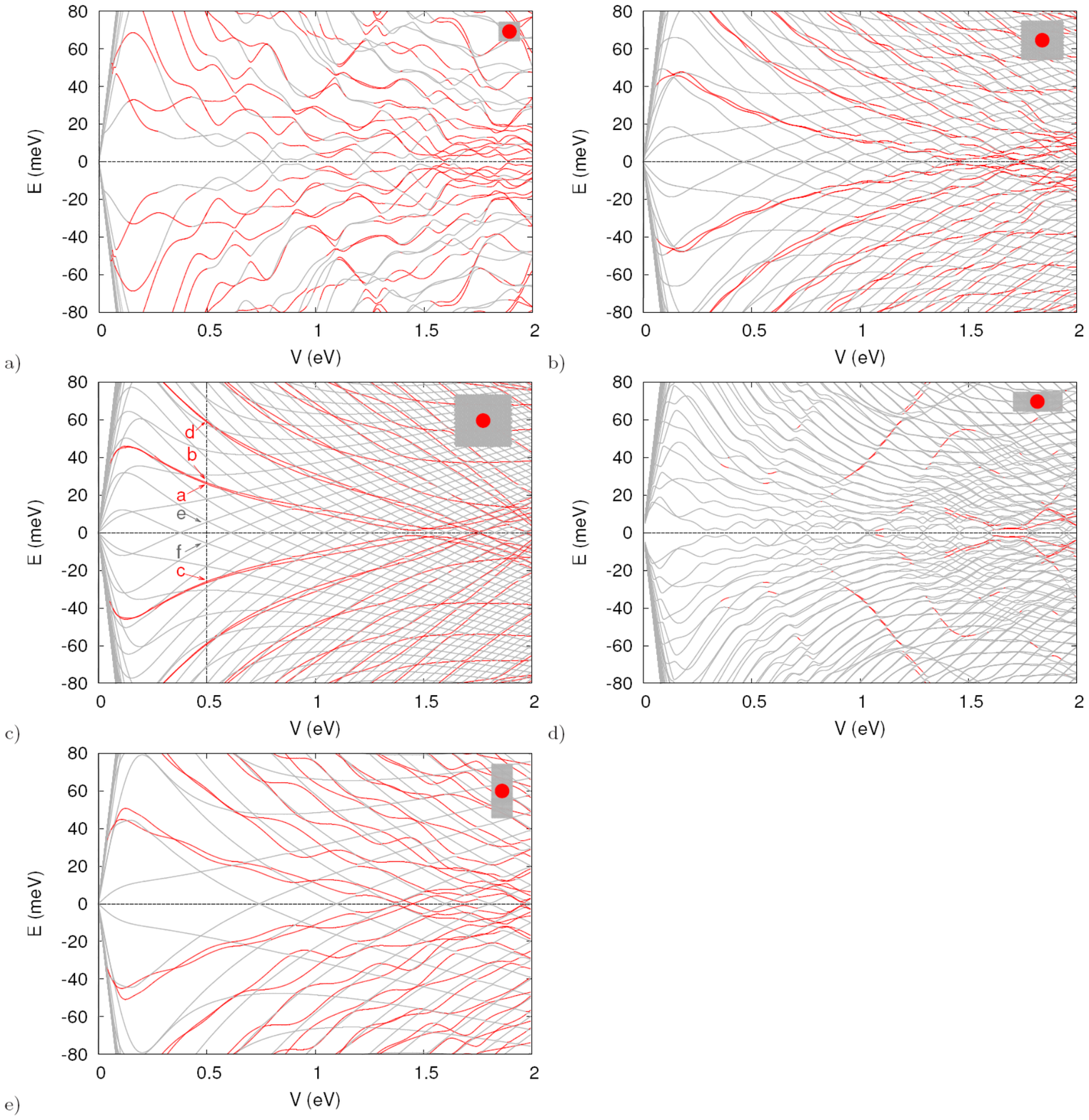}

\vspace{-8cm}
\caption{Energy spectrum for a dot of radius $R=5$ nm defined within the center of a rectangular flake
of Fig. \ref{schemat2} as a function of the potential step $V$  for
a) $N_z=60$, $N_{ar}=70$ (14.76 nm $\times$ 14.91 nm),
b) $N_z=118$, $N_{ar}=138$ (29.02 nm $\times$ 29.39 nm),
c) $N_z=166$, $N_{ar}=190$ (40.83 nm $\times$ 40.47 nm),
d)  $N_z=200$, $N_{ar}=70$ (49.19 nm $\times$ 14.91 nm),
e) $N_z=60$, $N_{ar}=200$  (14.76 nm $\times$ 42.6 nm).
With the red (grey) color we plot the energy levels of states which are localized within (outside) the dot.
The inset shows schematically the geometry the flake versus the size of the dot.
The letters in (c) indicate the energy levels whose wave functions are plotted in the corresponding panels
of Fig. \ref{ffu}.
}
\label{stanynaj}
\end{figure*}

\begin{figure}[ht!]
\hspace*{-1.6cm}
\includegraphics[scale=0.41,angle=-90]{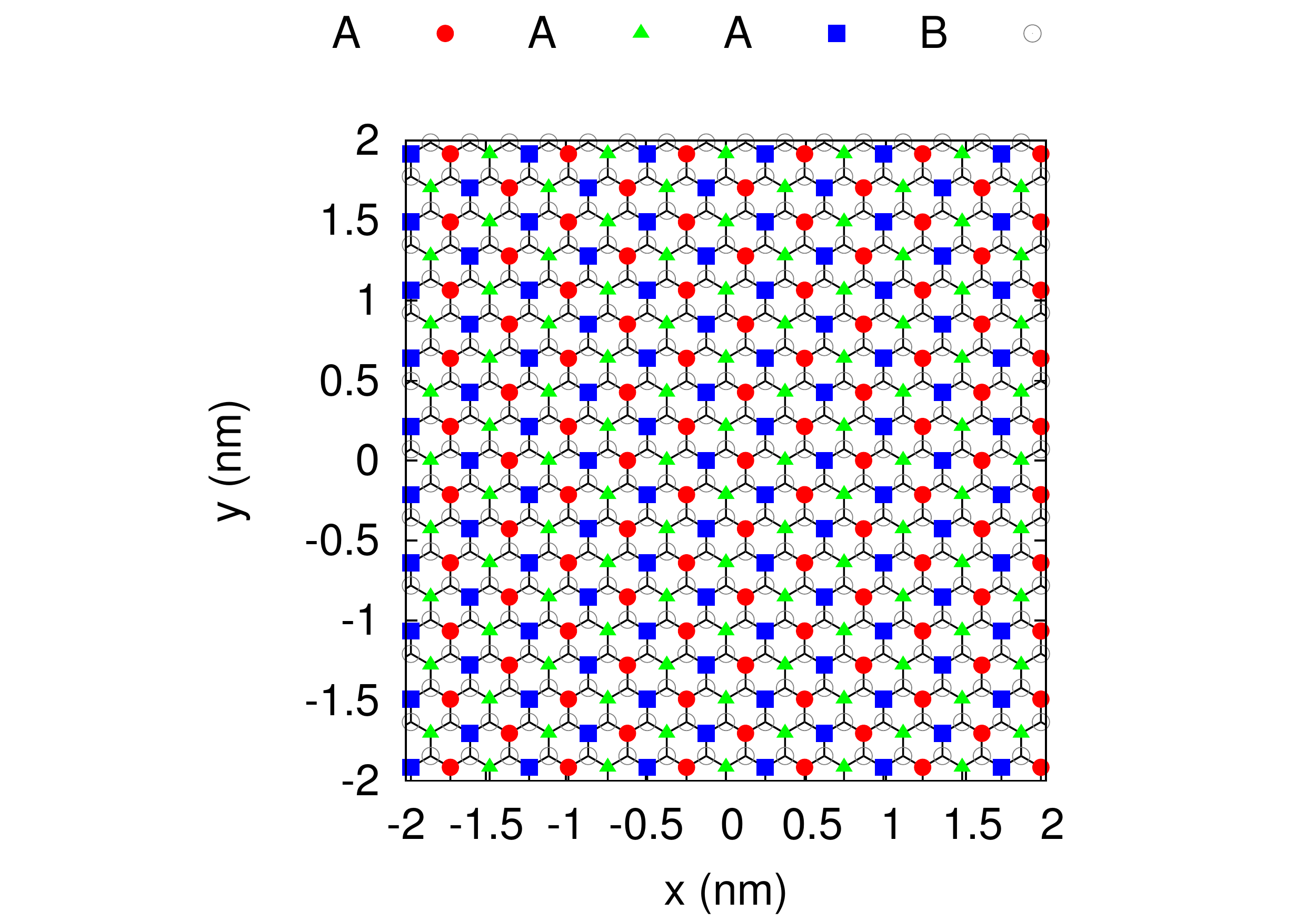}
\caption{A single layer of graphene. The open circles (full symbols) indicate B (A) sublattice.
Within A sublattice we select three columns of atoms on lines parallel to the armchair boundaries.
The selection is performed for discussion of wave functions of energy levels near the zero energy. }
\label{siatka}
\end{figure}

\begin{figure*}[ht!]
\begin{center}
\hspace*{-2.6cm}
\includegraphics[scale=1.05]{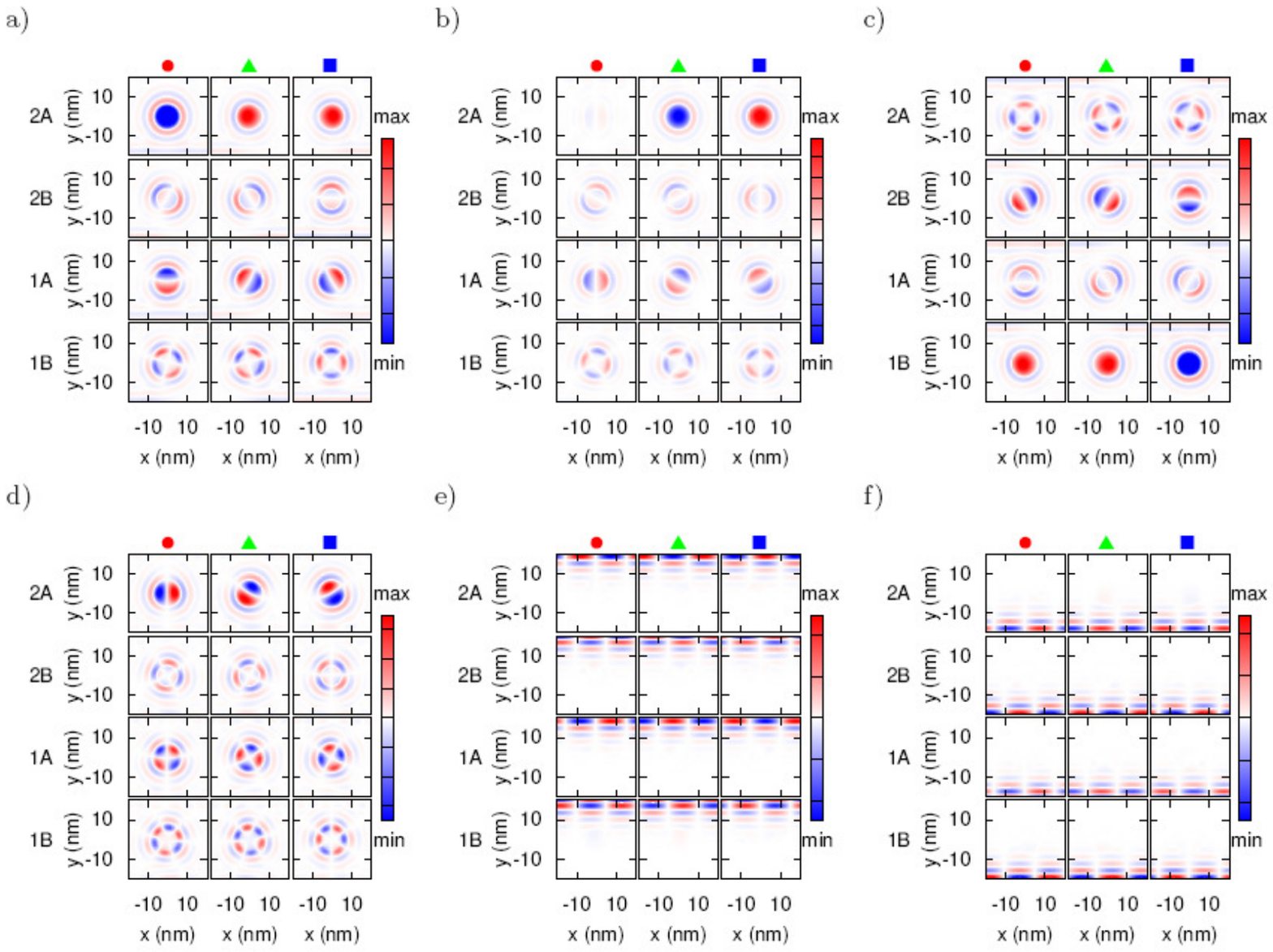}
\end{center}

\vspace{-16.7cm}
\caption{Real parts of wave functions of energy levels presented in Fig. \ref{stanynaj}(c) for $V=500$ meV. Pannels (a-f) correspond to energy levels
indicated by same letter in Fig. \ref{stanynaj}(c). In each plot the rows from top to bottom correspond to: sub-lattice A in the upper layer (2A),
sub-lattice B in the upper layer (2B), sublattice A in the lower layer (1A) and sublattice B in the lower layer (1B).
The columns in each of the plots correspond to different sub-sublattices selected as in Fig. \ref{siatka} to remove the rapid
oscillation from the plot.
For continuous approximation and rotational symmetry \cite{pereira} the components of wave functions
are of form $\cos((m-1)\phi),\cos((m)\phi), \cos((m)\phi), \cos((m+1)\phi)$ up to a phase for sublattices 2A, 2B, 1A, and 1B, respectively,
with $m$ -- an integer quantum number.
Plots a), b), c), and d) correspond to $m=1$, $m=1$, $m=-1$, and $m=2$, respectively.
For illustration a contribution from atomic site $i$ was taken with a finite spatial extension assumed with by a Gaussian $\exp(-|{\bf r}-{\bf r}_i|^2/\sigma^2)$,
where $\sigma=0.47$ nm is more or less the distance between the nearest atoms of the same sub-sublattice.
The color scale is the same in each of the plots.}
\label{ffu}
\end{figure*}

\begin{figure}[ht!]

\hspace*{-0.4cm}

\includegraphics[scale=0.34,angle=-90]{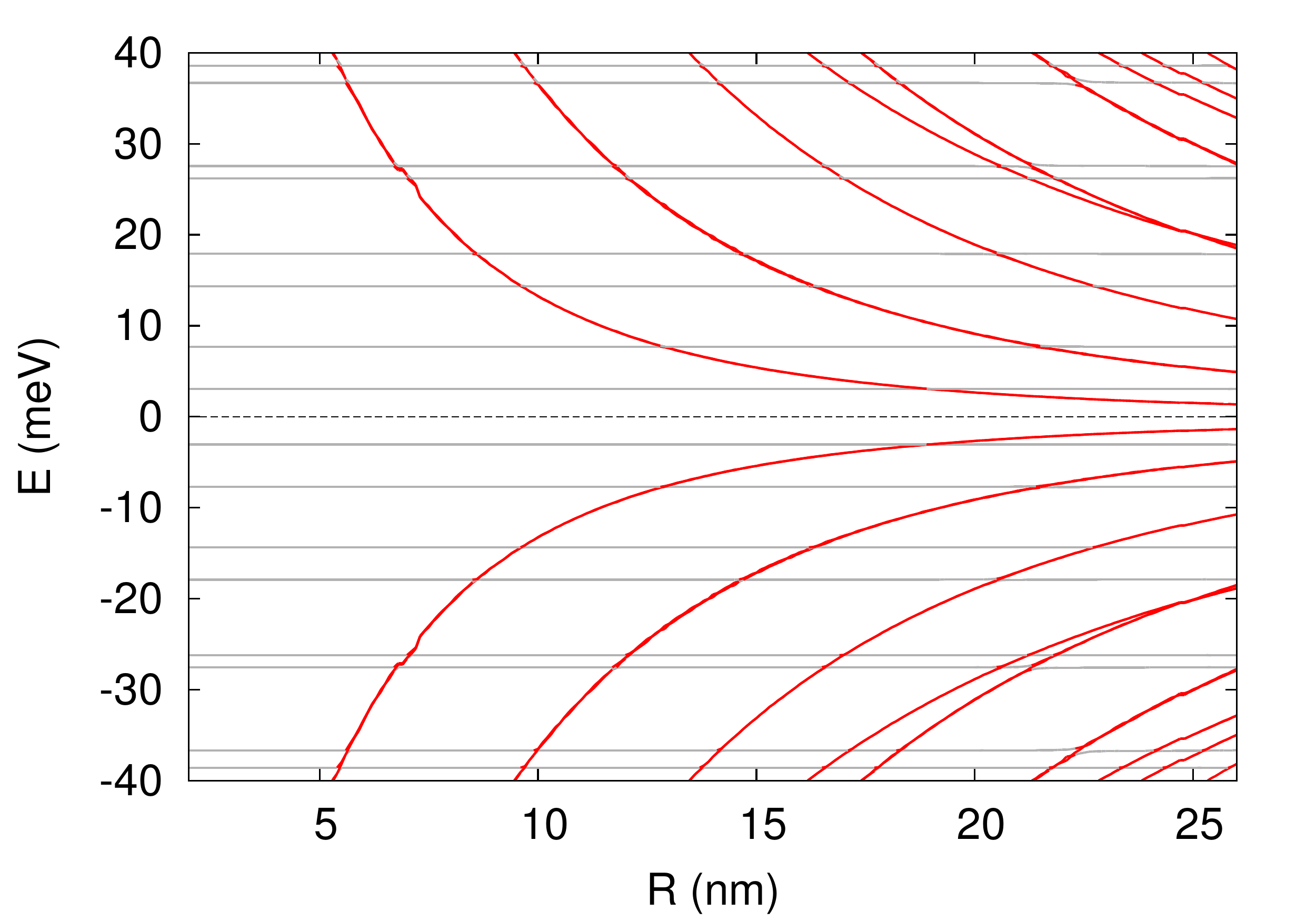}
\caption{
Energy spectrum for a dot of depth $V=200$ meV defined within the center of a large rectangular flake
($N_z=324$, $N_{ar}=380$ (79.69 nm $\times$ 80.94 nm) in function of dot radius. The red (gray) curves
correspond to energy levels inside (outside) the dot.
}
\label{stanyr}
\end{figure}

\begin{figure}[ht!]
\hspace*{-0.3cm}

\includegraphics[scale=0.3]{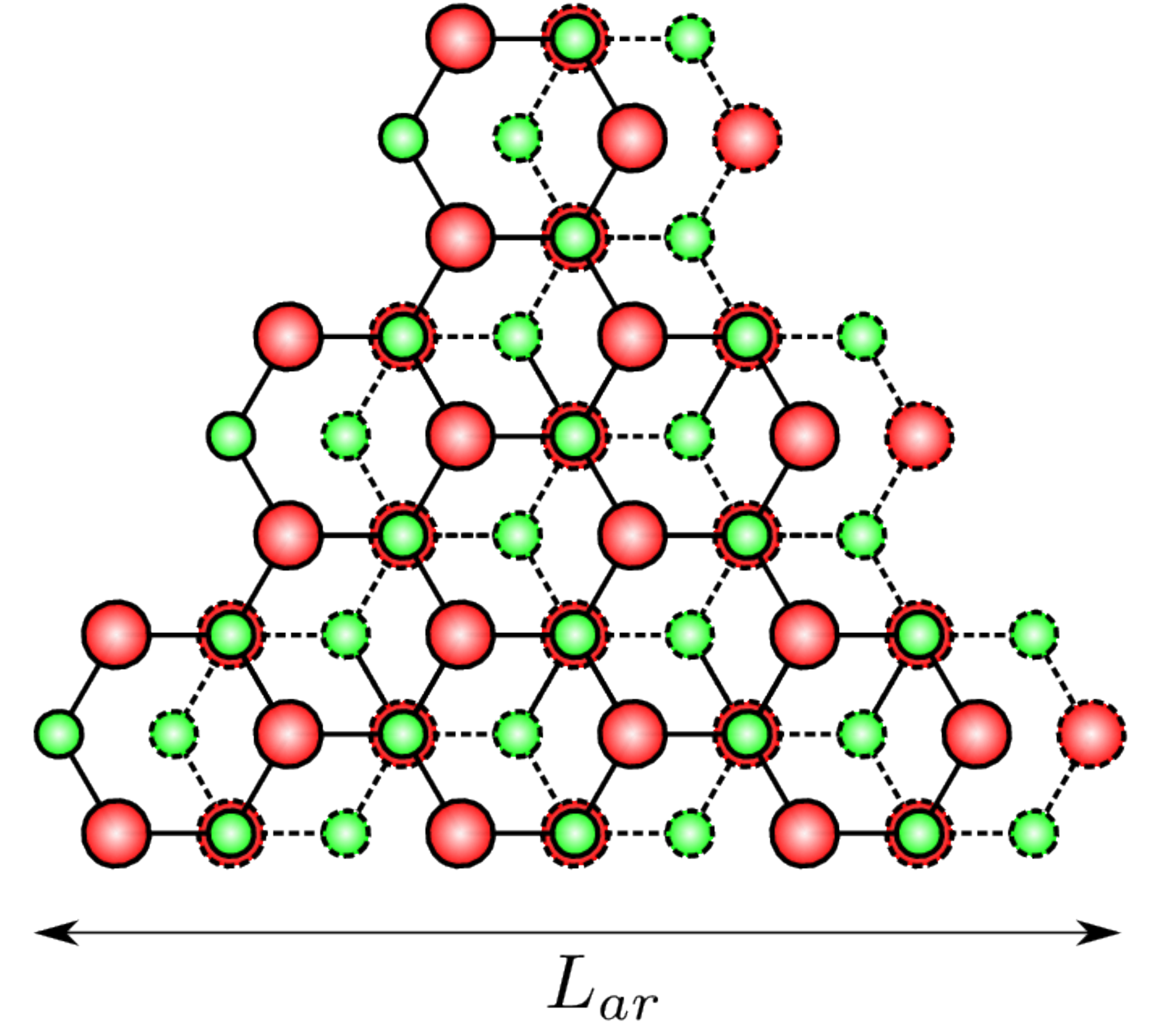}
\caption{Same as Fig. \ref{schemat2} only for a triangular flake with boundaries formed uniquely by armchair edges.
The two layers are shifted in the horizontal direction on the plot.}
\label{arm}
\end{figure}
\begin{figure}[ht!]
\hbox{
\hspace*{-0.4cm}
	   \begin{tabular}{c}
	   
           a) \includegraphics[scale=0.34]{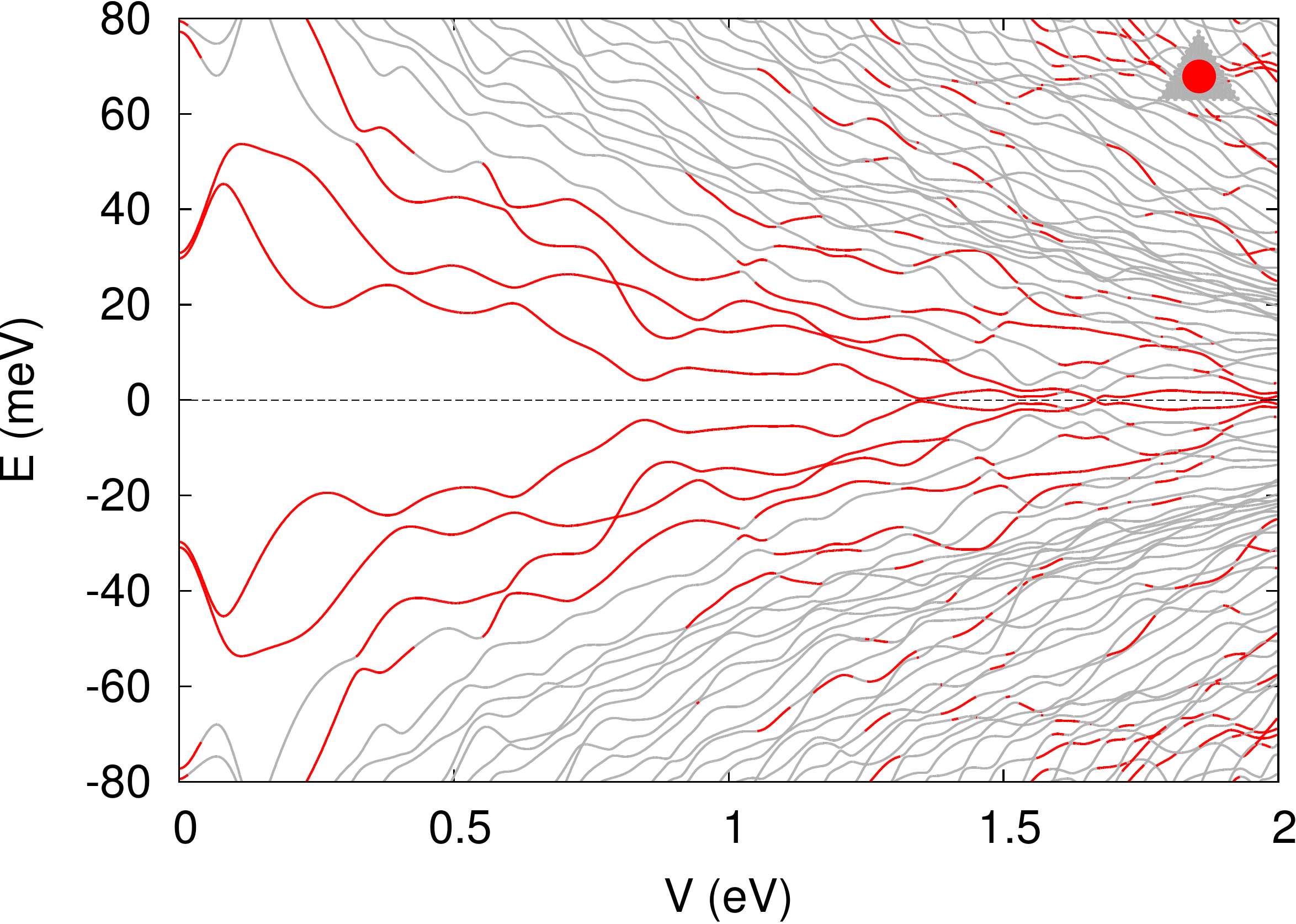} \\
b) \includegraphics[scale=0.34]{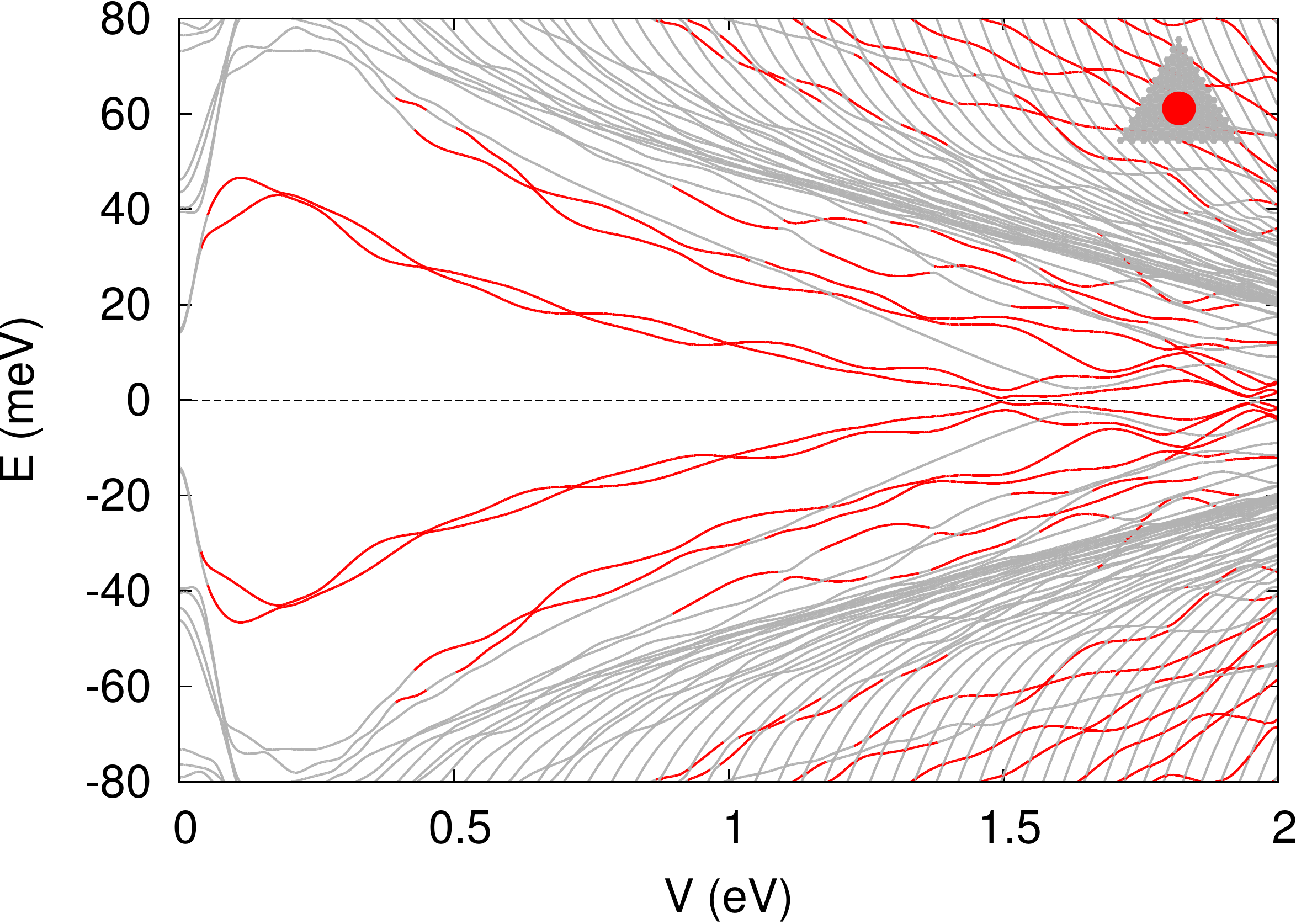} \\
c) \includegraphics[scale=0.34]{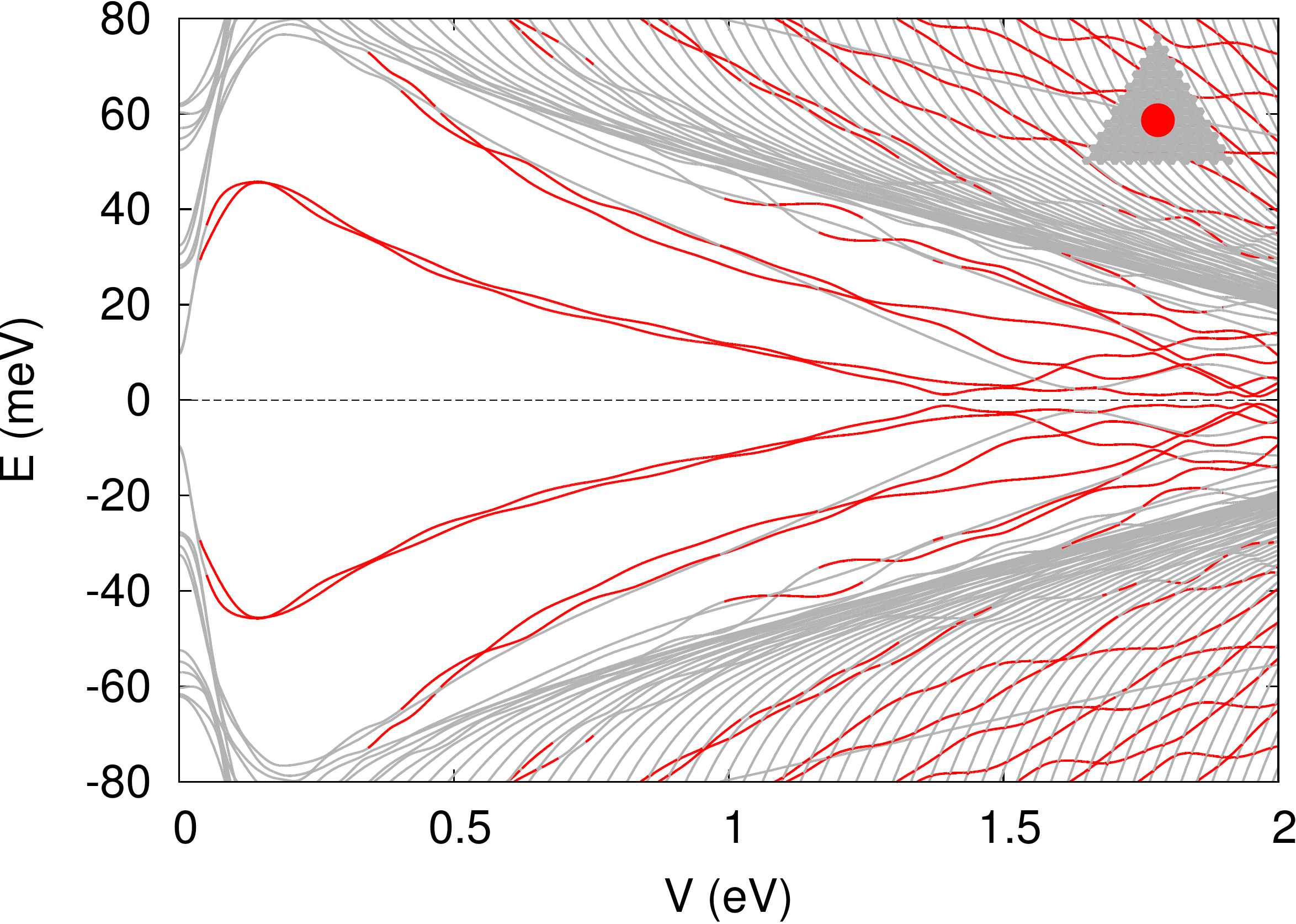} \end{tabular}}
\caption{Same as Fig. \ref{stanynaj} for a triangular flake [Fig. \ref{arm} with armchair edges only] with side length $L_{ar}=25.56$, $38.34$ and 46.86 nm, respectively,
and dot of radius $R=5$nm.}
\label{stanyt}
\end{figure}
\begin{figure}[ht!]
\hbox{
\includegraphics[scale=0.34,angle=-90] {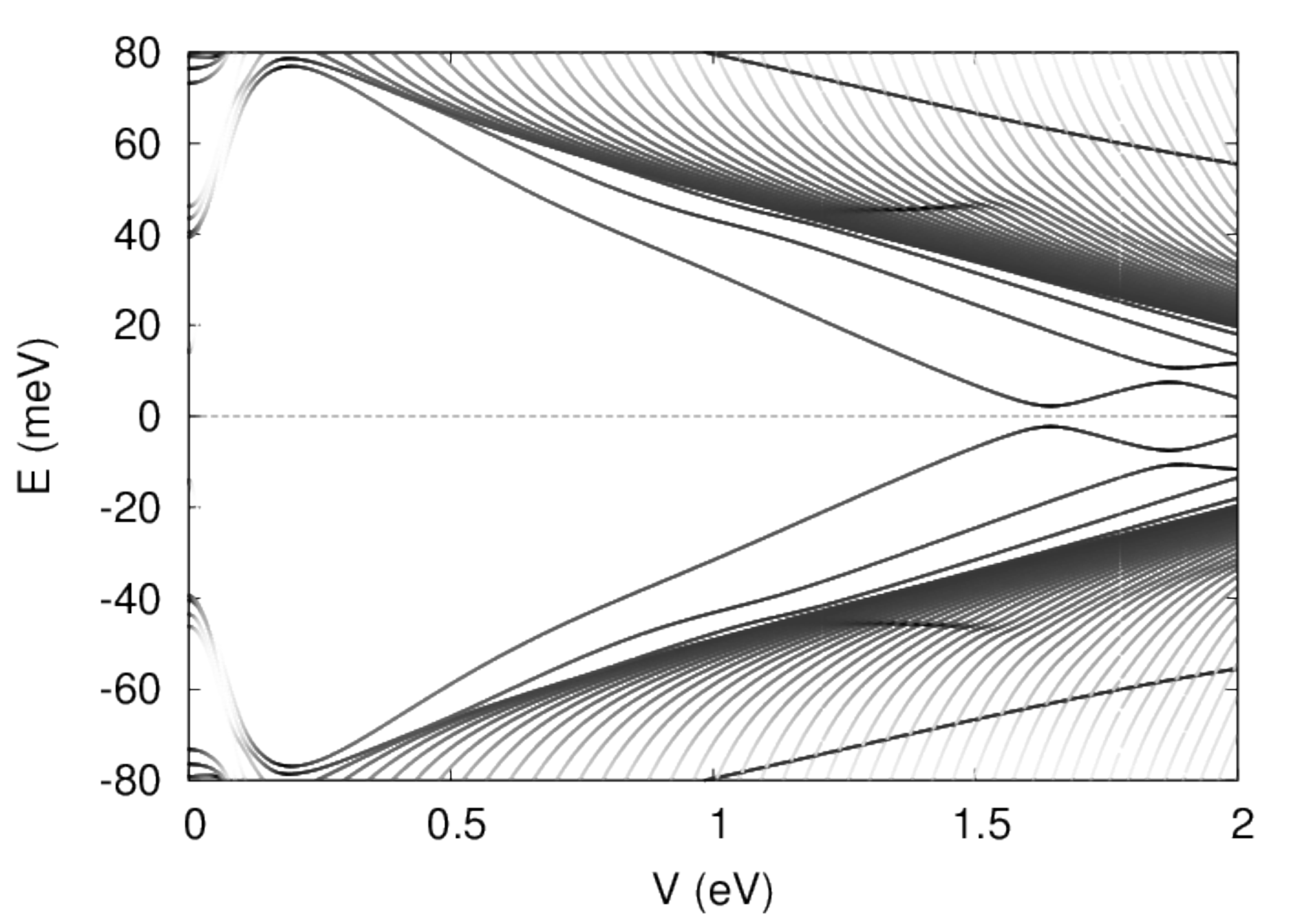} }
\caption{Same as Fig. \ref{stanyt}(b) only without the dot inside the flake ($R=0$).
The energy levels are plotted with varied shades of gray depending on the distribution
of the wave functions between the layers. We calculate the probability of finding
the electron on the two layers, next we calculate $c=(\frac{s}{l})^4$, where $s$ and $l$
stand for smaller and larger probabilities respectively. The value of $c=1$ ($c=0$) corresponds to the black (white) color.
}
\label{bezk}
\end{figure}

\section{Results and Discussion}

\subsection{Rectangular flake with zigzag and armchair edges}

The energy spectrum calculated for the dot of radius $R=5$ nm defined
on a rectangular flake of Fig. \ref{schemat2} is displayed in Fig. \ref{stanynaj}.
For $V=0$ a degenerate energy level of zero energy is observed.
This is a characteristic signature of the zigzag edges.\cite{gqd,cons,neg,peeters}
When $V$ increases,  energy levels which are totally localized at these edges
follow $E=\pm V$ dependence, with $+V$ ($-V$) on the dangling zigzag edge of the upper (lower) layer.
With the red (grey) color we plotted in Fig. \ref{stanynaj} the energy levels which are localized inside (outside) the dot.
In order to distinguish between the two, we calculate the probability of finding the electron closer to the dot center than $1.2 R$ and when it exceeds  60\% we conclude that the state is dot-localized.
For the smallest flake [Fig. \ref{stanynaj}(a)] localization of the energy level
changes several times for a single level as $V$ increases which indicates a strong coupling between the dot and its outside.
For the  flake of the side length equal to $4R$ [Fig. \ref{stanynaj}(c)] the dot-localized energy levels acquire a continuous identity.
Moreover, we can see that  the dot-localized states tend towards a two-fold degeneracy for larger size of the flake.
In the continuum approach with the
infinite graphene plane this
corresponds to the valley degeneracy\cite{grar}  which is lifted in the tight-binding
approach for a finite flake (see the discussion below).
The avoided crossings between the dot and the edge-localized states can still be observed in  [Fig. \ref{stanynaj}(c)]  which
indicates a presence of the tunnel coupling between the dot and the zigzag edge. Note,
that the distance between the dot and the edge is now as large as $3R$,
and still the coupling is observed. Figure \ref{stanynaj} shows that the density of the delocalized states near $E=0$ increases with
increasing size of the flake. A continuum of delocalized states should be expected for the infinite flake,
so that the dot-localized states appear within the continuum and are not separated by any energy gap
from the delocalized states in contrast to  quantum dots defined in bulk semiconductors.\cite{jacak}

\subsection{Symmetries of localized wave functions states}

Let us discuss the symmetries of wave functions and localization of energy levels for the large flake of Fig. \ref{stanynaj}(c).
The tight-binding method produces amplitudes of wave functions at the atomic sites. The wave functions near zero energy vary very fast from site to site,
which makes the analysis of wave functions a rather cumbersome task as compared to the continuum approach \cite{peeters} that produces a smooth envelope with the rapid variation
delegated to the Bloch functions. Nevertheless, we found that the symmetries of the smooth envelope can still be extracted from the tight-binding wave functions, since
the latter near zero energy turn out to oscillate on lines of atomic sites that are parallel to the armchair boundary. We select these lines as sub-sublattices
in a way that is explained in Fig. \ref{siatka}.

Figure \ref{ffu} shows the wave functions on the corresponding sub-sublattices for $V=0.5$ eV and 6 energy levels selected in Fig. \ref{stanynaj}(c).
Figures \ref{ffu}(a) and (b) correspond to the two lowest-energy nearly degenerate quantum-dot-localized energy levels [Fig. \ref{stanynaj}(c)].
 The corresponding components of both wave functions oscillate in function of the angle in the same way.
 In the continuum approach for an infinite graphene layer each of the confined energy levels is two-fold degenerate
 with respect to the valley.\cite{pereira} Each of the degenerate energy levels correspond to the same angular symmetries.\cite{pereira}
In the present results the valley degeneracy is lifted by the edge, but both the wave functions preserve the same symmetries.
In the continuum approach, for the low-energy effective Hamiltonian with circular confinement \cite{pereira} each of these wave function components
possess a definite angular momentum. Subsequent wave functions corresponding to 2A, 2B, 1A, 1B sublattices are
described by angular wave functions of form $\exp(i(m-1)\phi)$, $\exp(i m\phi)$, $\exp(im\phi)$,  $\exp(i(m+1)\phi)$, respectively, with
the same angular momentum in the sublattices forming the interlayer bonds 2B-1A.\cite{pereira}
We can see that the real parts of the wave functions components of Fig. \ref{ffu}(a,b) behave exactly as should be expected
for the quantum number $m=1$.  This behavior is quite remarkable taking into account that the circular external potential is
superposed on hexagonal lattices of the layers.
The form of the Dirac Hamiltonian of the continuum approach for rotationally symmetric potentials
implies that  $E(m)=-E(-m)$.
This feature is also found in the tight-binding solution -- see the plot for the energy level $c$
which is the inverse of energy level $a$, and its wave function [Fig. \ref{ffu}(c)] clearly indicates $m=-1$.
Note, that the dot-localized states for electrons ($E>0$), and for holes ($E<0$) are mainly localized within 2A and 1B sublattices,
respectively.

Figures \ref{ffu}(e) and (f) show the two wave functions of energy levels that are the closest to $E=0$ in Fig. \ref{stanynaj}(c) for $V=500$ meV.
We can see that they correspond to the zigzag edges -- the one which grows (decreases) in energy with the upper (lower)
zigzag edge.  These energy levels are evenly distributed on the two-layers of the flake / the four sublattices of Fig. \ref{ffu} and penetrate in the inside of the dot, hence
the avoided crossings observed in Fig. \ref{stanynaj}.

As presented above the wave functions for states of energies close to the Fermi energy
turn to change smoothly on three sub-sublattices selected as explained in Fig. \ref{siatka}.
This finding deserves a short comment.
For a single infinite plane of graphene one can attribute a definite
wave vector $k$ to the Hamiltonian eigenstates
\begin{equation}
 |{\psi^k}\rangle=\frac{1}{\sqrt{N}}\sum_{R} \exp{(i\vec{k} \vec{R} )} |{R}\rangle,
\end{equation}
where $|{R}\rangle$ stands for the electron $p_z$  orbital at the $R$ ion.
The wave function can be separated into components spanned by $A$ and $B$ sublattices 
\begin{equation} |{\psi^k}\rangle=|{\psi^k}\rangle_A+|{\psi^k}\rangle_B\end{equation} 
Consider next the $|{\psi^k}\rangle_A$ component.
All positions of the ions of the $A$ sublattice are generated by the lattice basis
vectors
\begin{equation}
 \vec{a_1}=(-\frac{\sqrt{3}}{2},\frac{3}{2})a \ \ \
  \vec{a_2}=(\frac{\sqrt{3}}{2},\frac{3}{2})a,
\end{equation}
$\vec{R_A}\equiv |n,m\rangle = n\vec{a_1}+m\vec{a_2}$, where  $n$ and  $m$ are integers.
The lowest energy states are associated with $K$ and $K'$ points at the Dirac
cones of the reciprocal space \cite{grar}
\begin{equation}
 \vec{K}=(\frac{4\pi}{3\sqrt{3}a},0)\ \ \
  \vec{K'}=-(\frac{4\pi}{3\sqrt{3}a},0).
\end{equation}
Inserting coordinates of the two valleys into the wave function
\begin{equation}
 |{\psi^{K(K')}}\rangle_A=\frac{1}{\sqrt{N}}\sum_{n,m} \exp{((-)\frac{2\pi i}{3}(m-n) )} |{n,m}\rangle.
\end{equation}
The amplitude which accounts for the ion $|n,m\rangle$ contribution to the wave function  $\exp{((-)\frac{2\pi i}{3}(m-n) )}$
takes on three different values, which are equal on each of the sub-sublattices depicted by color
symbols of Fig. \ref{siatka}.
In the present case -- with a finite size of the flake and modulated potential -- the wave vector is not a strictly good quantum number
and we obtain a smooth variation of the amplitudes instead their constant values on the sub-sublattices.

\subsection{Lifted valley degeneracy for dot-localized states}
The continuum low-energy approximation of the tight-binding Hamiltonian
for an infinite sheet of graphene \cite{pereira} predicts double degeneracy
of each dot-localized energy level with respect to valley.
   The results of Fig. \ref{stanynaj} indicate that:
a) the delocalized states in Fig. \ref{stanynaj} are not degenerate; b) the dot-localized energy levels
tend toward degeneracy only when the dot is defined within a larger flake.
 The possible reasons for lifting the valley degeneracy of the dot-localized energy levels
 are: i) proximity of the armchair edges which are known to couple the valleys in the single-layer graphene;\cite{brey}
ii) the zigzag dangling bonds: for a single-layer graphene the zigzag edges do not introduce the intervalley mixing,
but here on each of the two layers the flake terminates on atoms of a different sublattice [see Fig. \ref{schemat1}].
The contribution of ii) in lifting the degeneracy is quite evident in Fig. \ref{stanynaj}(c): see the lowest couple of dot-localized energy
levels near $V=0.25$ eV. Only one of these levels enters into an avoided crossing with  an energy level that is associated
with the zigzag edge, hence the lifting of the degeneracy within the regions of the avoided crossing, which
are wider for smaller flakes.
The significance of i) is not as evident: near $E=0$ no energy levels
associated with the armchair edge appear.

In order to explain the role of the separate types of edges in lifting the valley degeneracy of the dot-localized states we performed calculations for a strongly elongated rectangular flakes. We
kept one of the sides of the small flake of Fig. \ref{stanynaj}(a) and increased the other one. In this way one or the other type
of the edge becomes more strongly coupled with the dot localized states.
The results are displayed in Fig. \ref{stanynaj}(d) and Fig. \ref{stanynaj}(e), for flake dimensions $49.19$ nm $\times 14.91$ nm,
and $14.76$ nm $\times 42.6$ nm, respectively.
We can see that for the flake extended horizontally, the number of states associated with the zigzag edge in the considered energy range increases.
A larger number of zigzag states couples to the dot-localized ones, which are in this way destabilized due to the avoided crossings which now appear continuously with $V$.
The avoided crossings with the zigzag edge are distinctly limited when the edge is pushed further away from the dot: see Fig. \ref{stanynaj}(e).
In this case the distance to the zigzag edge is as large as in Fig. \ref{stanynaj}(c), only the armchair edge is closer.
We can see that the two-fold degeneracy of the dot-localized states is more distinctly lifted in Fig. \ref{stanynaj}(e) as
compared with Fig. \ref{stanynaj}(c). This is an evidence that proximity of the armchair edges also contributes to lifting
the valley degeneracy of the dot-localized states.

\subsection{Spectrum versus dot radius ($R$) for a large rectangular flake}

Let us now consider a very large flake of a side length of 80 nm and consider the energy spectrum in function of the dot size (see Fig. \ref{stanyr}).
We find that: i) the energies of dot-localized states decrease as $1/R^2$ in agreement with the Dirac calculations \cite{pereira}; ii) the states localized outside the dot -- which
in this energy range corresponds to the states localized at the zigzag dangling edges -- are completely insensitive to the size of the dot;
iii) the avoided crossings between the dot and the edge are not resolved at this scale;
iv) the valley degeneracy of the dot-localized states is nearly exact at this size of the flake.
We find that the dot-localized valley-degenerate wave functions near $R=20$ nm very well follow the $m$ quantum number description.
The dot-localized energy levels for $E>0$ follow the sequence: $m=1,2,3,0,4,5,-1,1$ in the growing energy order.

\subsection{Triangular flake with armchair edges only}

In quest for the energy gap between the dot-localized and delocalized states
we considered a flake with no zigzag boundaries. A triangular flake with armchair edges like the one
of Fig. \ref{arm} was assumed. The two layers were shifted in the horizontal direction,
so that both the layers include an equal number of dangling atoms. This allows
for a perfect electron-hole symmetry in the spectrum. For a general shift
this is not necessarily the case.
The energy spectrum calculated for triangular flakes is displayed in Fig. \ref{stanyt}.
We find a number of dot-localized states which are separated from the delocalized ones by a distinct energy gap.
However, the number of dot-localized states is  rather low -- in Fig. \ref{stanyt}(c) we find 4 levels at maximum (the number depends on $V$).
Including the spin degree of freedom, at most 8 first electrons added to the system will occupy the quantum dot.
For larger electron numbers the external states or the dot-localized ones will become occupied alternately.
Surprisingly, the energy threshold for the ionization of the electron
from the ground state is a non-monotonic function of the potential step $V$.
Moreover, the energy edge of the external states does {\it not} increase with the size of the triangle --
see Fig. \ref{stanyt}(b) and (c) for $V=2$ eV.

In order to inspect the peculiar properties of the energy gap in function of the interlayer voltage drop
we considered the case of $R=0$ (no dot) within the triangular flake, with a constant $\pm V$ potential
on both the layers. The energy levels are depicted in Fig. \ref{bezk}.
The six energy levels closest to the neutrality point $E=0$ are localized at the upper corner
of the triangle -- see Fig. \ref{arm}. The dense series of energy levels that are observed in Fig. \ref{bezk} and Fig. \ref{stanyt} further
away from $E=0$ are localized at the left and right edges of the triangle of Fig. \ref{arm}, where the
dangling bonds appear due to the horizontal shift of the layers.
The energy levels for $E<0$ ($E>0$) are localized at the left (right) edge where the upper (lower) layer is dangling.
The wave functions of all the three layers are almost equally distributed between the layers but
for $E<0$ ($E>0$) most of the wave functions stick to the upper (lower) -- dangling edge. Since the potential at the upper (lower) edge
increases (decreases) with $V$, we obtain the energy increase and hence the lowering of the gap with the increasing bias.

In Fig. \ref{stanyt} we notice, that the valley degeneracy of the dot-localized states is restored when the size of the triangle increases.
Note, that now we have uniquely the armchair edges, so the valley mixing
by this edge\cite{brey} is fully responsible for the lifted degeneracy.
Moreover -- when we compare Fig. \ref{stanyt}(c) and Fig. \ref{stanynaj}(c) -- for large flakes of rectangular and triangular
geometries -- we can see that the dot-localized states form the same pattern. Only, the avoided crossings
with states extending from the zigzag edges -- visible for the rectangular flake -- are no longer present for the triangular flake.

\subsection{Vacancies versus the dot-localized energy levels}
\begin{figure}[ht!]
\hbox{
	   \begin{tabular}{c}
           a)\includegraphics[scale=0.34,angle=-90]{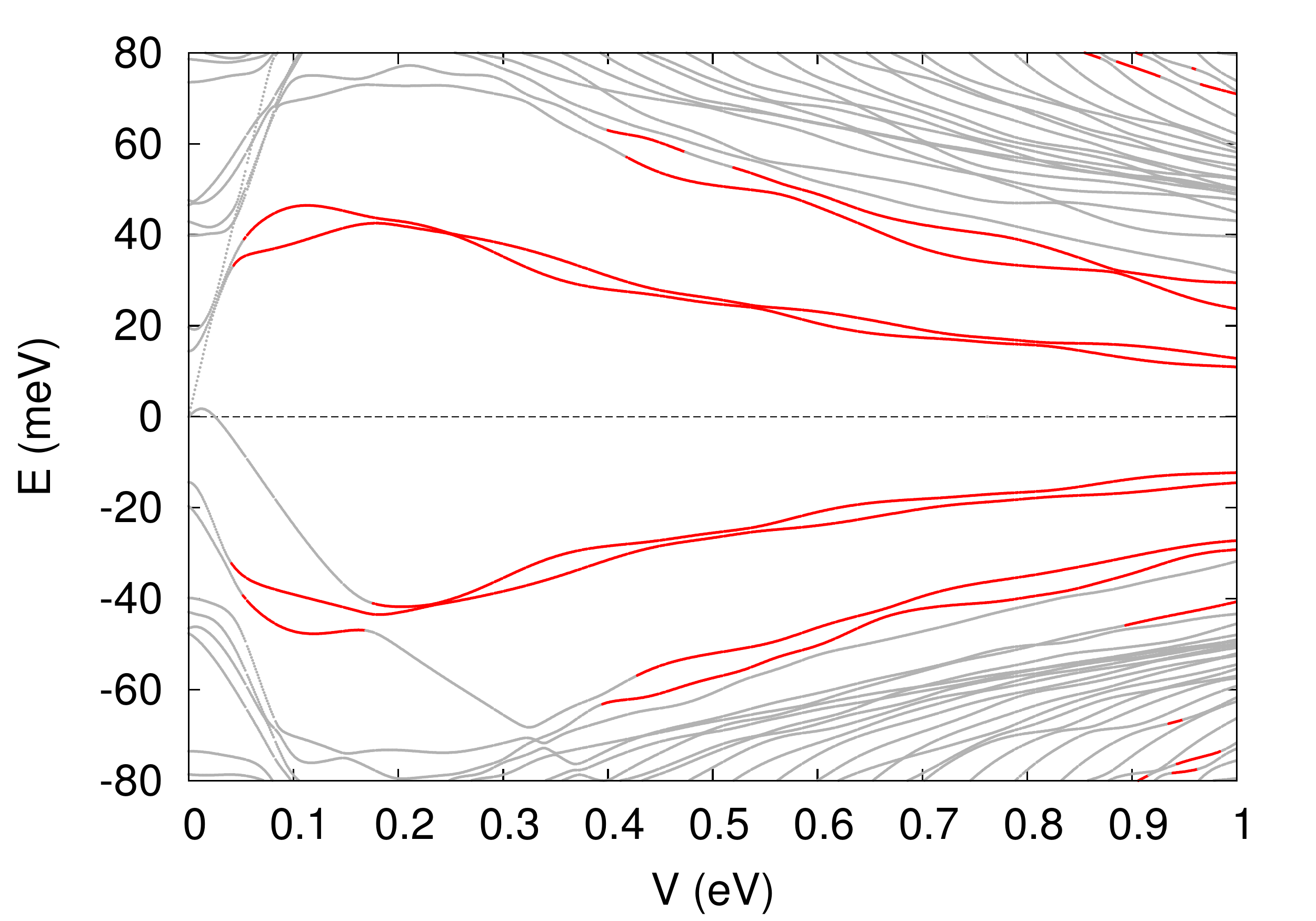} \\
b)\includegraphics[scale=0.34,angle=-90]{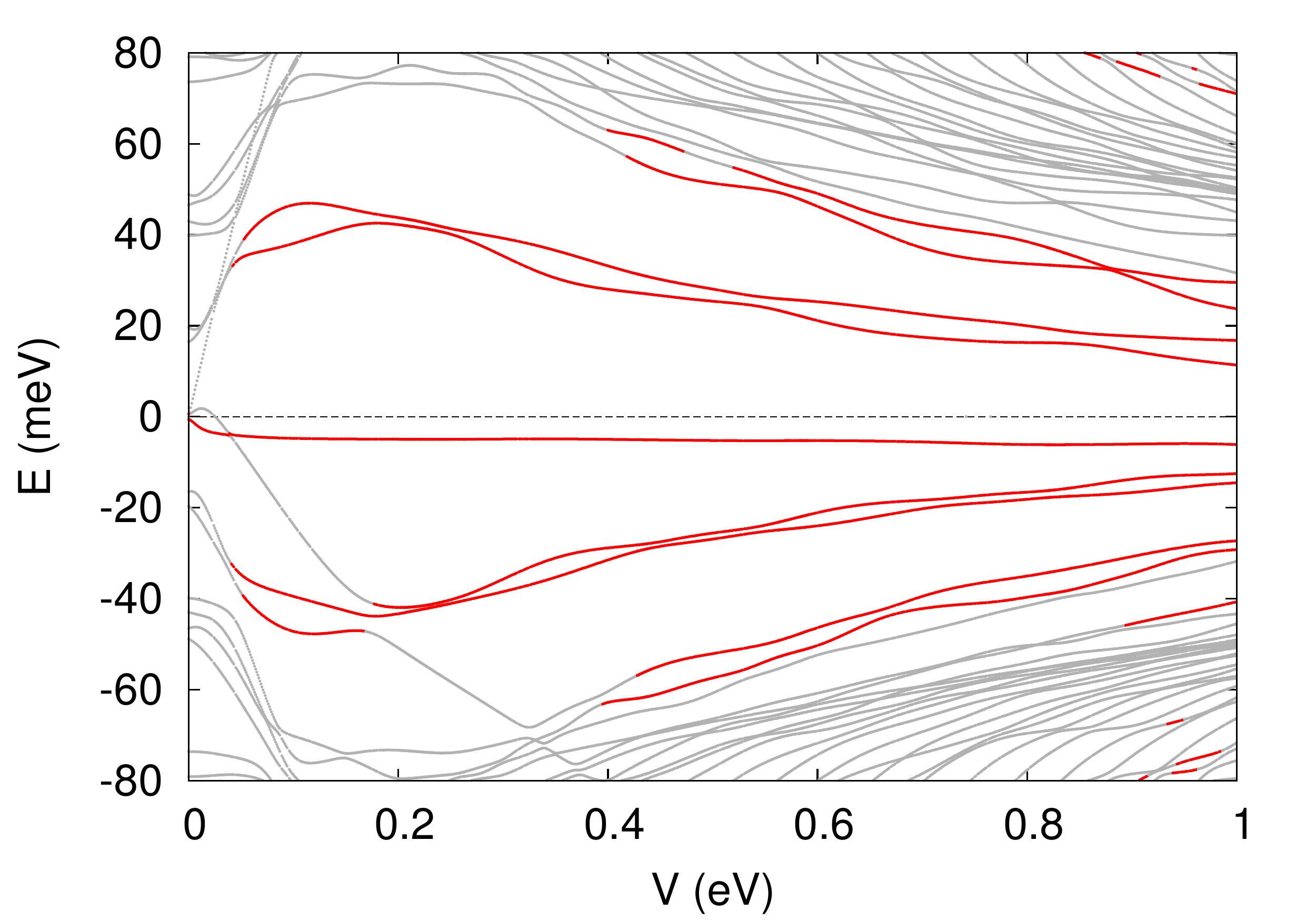} \end{tabular}}
\caption{Same as Fig. 8(b) only with (a) a single vacancy on both the layers outside the dot (b) and with the third
vacancy introduced to the lower layer inside the dot.}
\label{vacan}
\end{figure}

We have seen that the zigzag edges mix the dot-localized states with
the states in the rest of the flake. A natural question arises whether the  atomic size defects would introduce a similar
mixing. In order to answer this question we introduced the vacancies to the system considered in the
triangular flake of Fig. 8(b) with armchair external edges.
The spectrum with a single vacancy introduced to each of the layers outside the dot is displayed in Fig. \ref{vacan}(a).
The vacancies introduce two energy levels at $E=0$ in the absence of the bias. 
We assumed a missing atom of the 2B sublattice of the upper layer and a missing atom of the 1B sublattice of the lower layer.
The vacancy energy level of the lower layer
which falls for increasing $V$ enters into an avoided crossing with the dot localized energy level near $V\simeq 0.2$ eV and thus lifts
the electron-hole symmetry. The energy level associated with removed 2B sublattice atom changes much strongly with $V$, since
the vacancy lifts the interlayer coupling. 
We can see that in contrast to the zig-zag edge the vacancies do not form a quasi-continuous set of delocalized energy levels. In Fig. \ref{vacan}(b) we introduced a third vacancy inside the dot area in the lower graphene layer.
This vacancy gives rise to an extra energy level $E\simeq -4$ meV localized inside the dot with no further consequences to the rest
of the spectrum.

\subsection{Absence of the interlayer coupling}
\begin{figure}[ht!]
\hbox{
\includegraphics[scale=0.34,angle=-90] {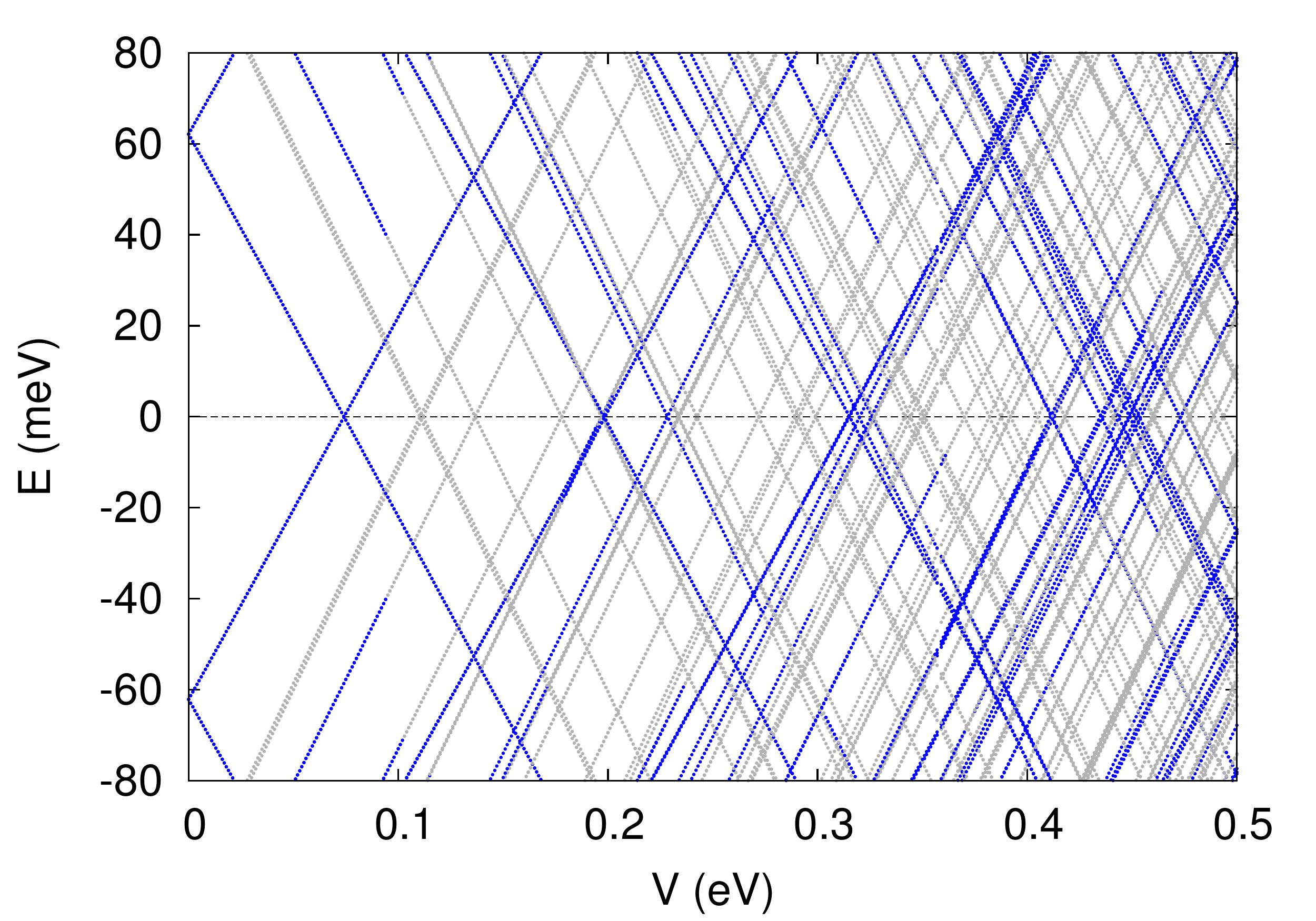}
}
\caption{Same as Fig. 8(b) only with neglected coupling between the graphene layers.
The blue (black) energy levels correspond to states in the lower (upper) layer.}
\label{brakt}
\end{figure}

The results presented above indicate that for large bilayer flakes containing zigzag edges
the dot-localized and extended states appear in the same energy range.
It does not imply that the idea for electrostatic confinement in bilayer graphene
does not work. In Fig. \ref{brakt} we showed the energy spectrum of the system of Fig. 8(b)
with neglected interlayer coupling. We have now a spatial modulation of the potential
with the quantum well in the upper layer and a central antidot region in the lower layer.
We have found that all the energy levels in the range shown in Fig. \ref{brakt} are localized
outside both the well and the antidot, hence the linear behavior of the energy levels with $V$.
This figure indicates no confinement induced by the potentia modulation within a single layer
of graphene, a fact which is usually attributed to the Klein theorem. 
Comparing Fig. \ref{brakt} with the energy spectra given above we can see that
the bilayer system with interlayer coupling supports the bound states inside the electrostatic
dot even when the bound states appear within the quasi continuum of delocalized energy levels.

\subsection{Flakes with reconstructed edges}
\begin{figure}[ht!]
\hbox{
\includegraphics[scale=0.3]{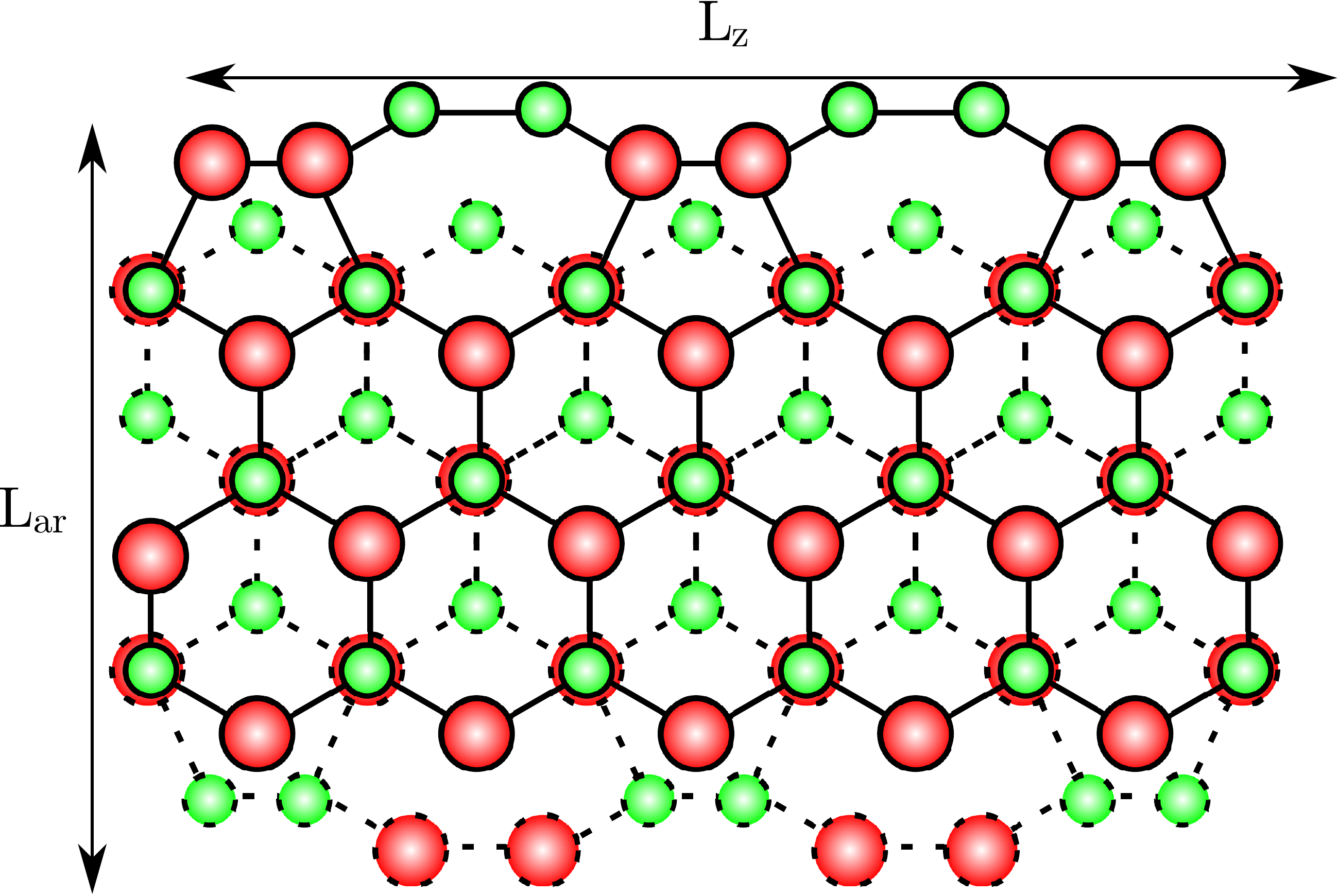}
          }
\caption{Same as Fig. \ref{schemat2} only with reconstructed edges replacing the zigzag ones.}
\label{reczag}
\end{figure}

\begin{figure}[ht!]
\includegraphics[scale=0.34,angle=-90]{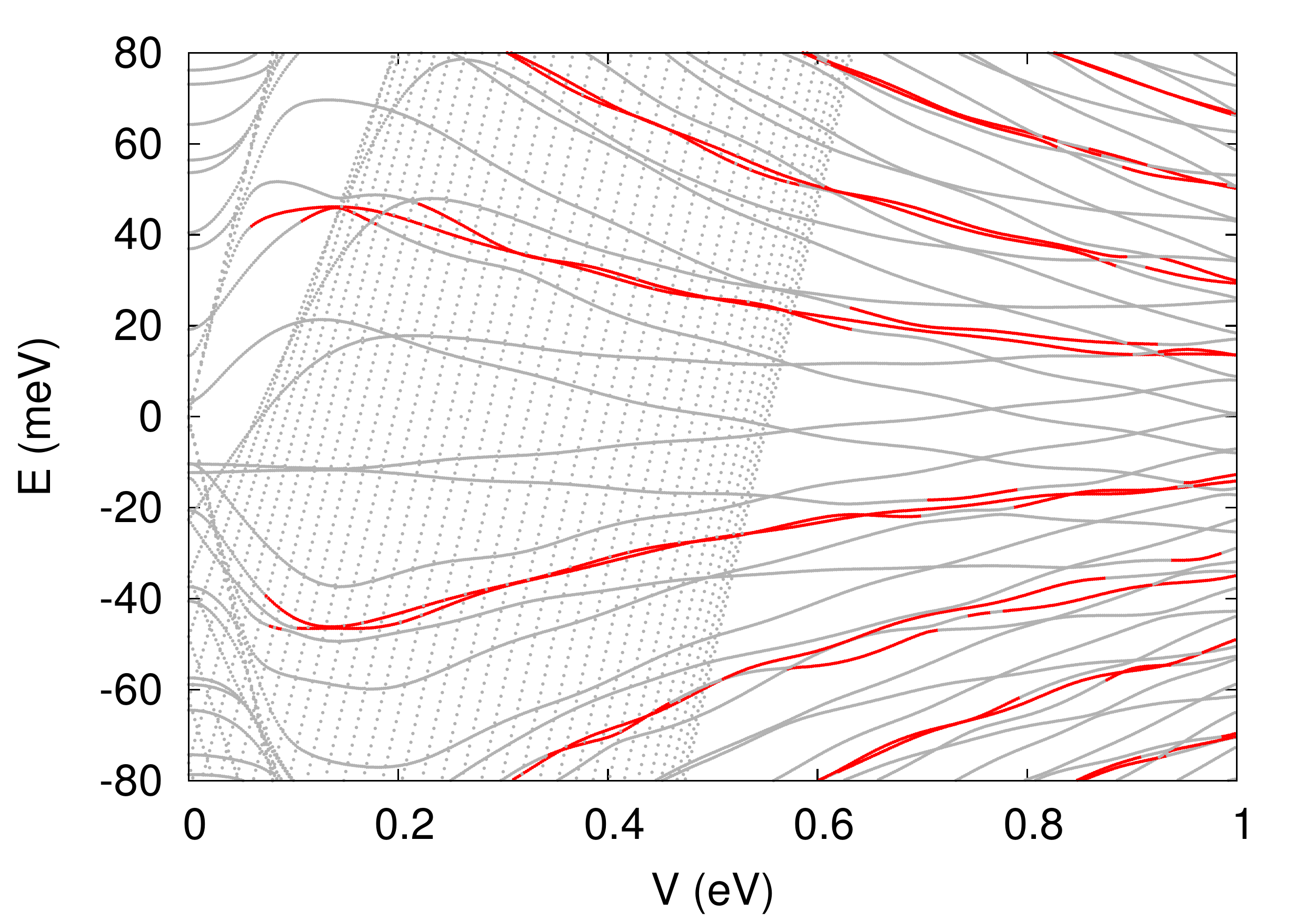}
\caption{Energy spectrum for a dot of radius $R=5$ nm defined within the center of a rectangular flake
with reczag edges
of Fig. \ref{reczag} as a function of the potential step $V$  for
$N_z=117$, $N_{ar}=138$ ($L_z=28.77$ nm, $L_{ar}=29.39$ nm. With the red (grey) color we plot the energy levels of states which are localized within (outside) the dot.
For the corresponding spectrum with hexagonal zigzag edge see: Fig. \ref{stanynaj}(b).
}
\label{recwi}
\end{figure}

Reference \onlinecite{koskinen} indicated that the hexagonal structure of graphene is unstable near
the edges with respect to formation of so called zz(57)  or reczag edge with pentagons and heptagons placed  alternatively along the boundary of the flake.
We considered a rectangular flake that is depicted in Fig. \ref{reczag} with dangling reczag edges replacing the zigzag edges
of Fig. \ref{schemat2}. The tight-binding parameters for the carbon atoms entering the pentagon and heptagon cells were
determined in Ref. \onlinecite{ost}. We use these parameters for our calculation. The calculated spectrum for the dot of radius $R=5$ nm, $L_z=28.77$ nm, and $L_{ar}=29.39$ nm
is presented in Fig. \ref{recwi}.
The energy levels corresponding to the reconstructed edge at $V=0$ appear degenerate below the neutrality point.
As $V$ is introduced the degeneracy is lifted. We observe a series of energy levels which increase linearly with $V$.
These energy levels correspond to the reczag edge on the upper layer of the flake.
Besides this series the spectrum is qualitatively similar to the case of the zigzag edges of Fig. \ref{stanynaj}(b)
with the exception of the electron-hole asymmetry which is introduced by the reczag edge.

\section{Summary and Conclusions}
 We performed a tight-binding analysis of quantum dots defined by external potentials
within bilayer graphene flakes of a finite size. We discussed localization of charge
carriers within the dot, the coupling of the dot to the zigzag and armchair edges
and the energy separation between the dot-localized states and the states localized outside the dot.
We extracted smooth envelope wave functions from the tight-binding eigenstates near the Fermi energy,
which allowed us to discuss the angular symmetries of the confined states. We found a close
correspondence to the symmetry properties of the continuous Dirac approximation of the tight-binding Hamiltonian.

In presence of the zigzag boundary we found a
number of dot-localized energy levels that are crossed or avoided crossed by energy levels of states extending from the zigzag edges.
The latter distinctly tend to form a continuum for large flakes of bilayer graphene. In this case the carriers
can be removed from the dot at no energy expense. For finite flakes the states outside the dot are discrete, and the
electrons when added to the system in the charging experiments will occupy alternatively the dot or the external area of the flake.
For a flake with uniquely armchair edges we found that the dot-localized energy levels appear within an energy window
which is inaccessible to the external states. In this case the graphene dot will possess the properties
similar to the conventional quantum dots of bulk semiconductors, i.e. a finite energy will be required
to remove the electron or the hole from the dot. Surprisingly, the energy window for the dot-localized states decreases at large depths
of the confinement potential.

We discussed the valley degeneracy of the dot-localized state which is only found for large distance
between the dot and the edges of the flake.
We found that the proximity of the armchair edges lifts the valley degeneracy of dot-localized states in a manner independent of the depth of confinement potential.
On the other hand the states extending from the zigzag boundaries -- which are not valley degenerate --
lift the valley degeneracy of the confined energy levels entering into avoided crossings with one of the pairs of the dot-localized energy levels.

\section*{Acknowledgments}
This work was supported by
PL-Grid Infrastructure. Calculations were performed in ACK--
CYFRONET--AGH on the RackServer Zeus.


\begin{thebibliography}{99}
\bibitem{grar} A.H. Castro Neto, F. Guinea, N.M.R. Peres, K.S. Novoselov, and A.K. Geim, Rev. Mod. Phys. {\bf 81}, 109 (2009).
\bibitem{gqd} P.G. Silvestrov and K.B. Efetov, Phys. Rev. Lett. {\bf 98}, 016802 (2007); B. Wunsch, T. Stauber, and F. Guinea,
Phys. Rev. B {\bf 77}, 035316 (2008); A. Matulis and F.M. Peeters, Phys. Rev. B {\bf 77}, 115423 (2008); A. R. Akhmerov and C. W. J. Beenakker, Phys. Rev. B {\bf 77},
085423 (2008); Z. Z. Zhang, K. Chang, and F. M. Peeters, Phys. Rev. B {\bf  77},
235411 (2008); J. Wurm, A. Rycerz, \ifmmode \dot{I}\else \.{I}\fi{}. Adagideli, M. Wimmer, K. Richter, and H.U. Baranger, Phys. Rev. Lett. {\bf 102}, 056806 (2009); A. G\"u\ifmmode \mbox{\c{c}}\else \c{c}\fi{}l\"u, P. Potasz, O. Voznyy,
M. Korkusinski, and P. Hawrylak, Phys. Rev. Lett {\bf 103}, 246805 (2009); F. Libisch, C. Stampfer, and J. Burgd\"orfer, Phys. Rev. B
79, 115423 (2009);  S. Schnez, F. Molitor, C. Stampfer, J. G\"uttinger, I. Shorubalko, T. Ihn, and K. Ensslin, Appl. Phys. Lett. {\bf 94}, 012107 (2009).
\bibitem{cons}J. G\"uttinger, T. Frey, C. Stampfer, T. Ihn, and K. Ensslin, Phys. Rev. Lett. {\bf 105}, 116801 (2010).
\bibitem{neg} B. Trauzettel, D.V. Bulaev, D. Loss, and G. Burkard, Nat. Phys. {\bf 3}, 192 (2007).
\bibitem{walas} P. R. Wallace, Phys. Rev. {\bf 71}, 622 (1947).
\bibitem{klein} O. Klein, Z. Phys. {\bf 53}, 157 (1929);  M. I. Katsnelson, K. S. Novoselov, and A. K. Geim, Nat. Phys. {\bf 2} 620 (2006);
J. M. Pereira Jr., F. M. Peeters, A. Chaves, and G. A. Farias, Semicond. Sci. Technol. {\bf 25}, 033002 (2010).
\bibitem{uwaga} J.H. Bardarson, M. Titov, and P.W. Brouwer, Phys. Rev. Lett. {\bf 102}, 226803 (2009).
\bibitem{peeters}  M. Zarenia, A. Chaves, G. A. Farias, and F. M. Peeters, Phys. Rev. B {\bf 84}, 245403 (2011).
\bibitem{jaskolski} W. Jask\'olski, A. Ayuela, M. Pelc, H. Santos and L. Chico, Phys. Rev. B {\bf 83}, 235424 (2011).
\bibitem{potasz} P. Potasz, A.D. G\"u\ifmmode \mbox{\c{c}}\else \c{c}\fi{}l\"u, and P. Hawrylak, Phys. Rev. B {\bf 81}, 033403 (2010).
\bibitem{wurmetal} J. Wurm, K. Richter and \ifmmode \dot{I}\else \.{I}\fi{}. Adagideli, Phys. Rev. B {\bf 84}, 075468 (2011).
\bibitem{review}   E. McCann and M. Koshino, Rep. Prog. Phys. {\bf 76}, 056503 (2013).
\bibitem{gap} E. McCann, Phys. Rev. B {\bf 74}, R161403 (2006); E. V. Castro, K. S. Novoselov, S. V. Morozov, N. M. R. Peres, J. M. B. Lopes dos Santos, J. Nilsson, F. Guinea, A. K. Geim, and A. H. Castro Neto,
Phys. Rev. Lett. {\bf 99}, 216802 (2007);  F. Xia, D.B. Farmer, Y.M. Lin, P. Avouris, Nano. Lett. {\bf 10}, 715 (2010); J. Park, S.B. Jo, Y.-J. Yu, Y. Kim, J.W. Yang, W.H. Lee, H.H. Kim, B.H. Hong, P. Kim, K. Cho, and K.S. Kim, Adv. Matt. {\bf 24}, 407 (2012).
\bibitem{pereira} J.M. Pereira Jr., P. Vasilopoulos, and F.M Peeters, Nano. Lett. {\bf 7}, 946 (2007).
\bibitem{pereirar} L. J. P. Xavier, J. M. Pereira, A. Chaves, G.A. Farias, and F. M. Peeters, Appl. Phys. Lett. {\bf 96}, 212108 (2010).
\bibitem{dro} S. Dr\"oscher, J. G\"uttinger, T. Mathis, B. Batlogg, T. Ihn, and K. Ensslin, Appl. Phys. Lett. {\bf 101}, 043107 (2012).
\bibitem{goo} A. M. Goossens, S.C.M. Driessen, T.A. Baart, K. Watanabe, T. Taniguchi, and L.M.K.  Vandersypen, Nano. Lett. {\bf 12}, 4656 (2012).
\bibitem{yac} M.T. Allen, J. Martin, and A. Yacoby, Nature Comm. {\bf 3}, 934 (2012).
\bibitem{els} L. Huang, Y.-C. Lai,and C. Grebogi, Phys.
Rev. E {\bf 81}, R055203 (2010); A. Rycerz, Phys. Rev. B {\bf 85}, 245424 (2012).
\bibitem{dbgb} J.B. Oostinga, H.B. Heersche, X. Liu, A.F. Morpurgo, and L.M.K. Vandersypen, Nature Mat. {\bf 7}, 151 (2008);
T. Taychatanapat, and P. Jarillo-Herrero, Phys. Rev. Lett. {\bf 105}, 166601 (2010); R.T. Weitz, M.T. Allen, B.E. Feldman, J. Martin, A. Yacoby,
Science {\bf 330}, 812 (2010);
J.D. Sanchez-Yamagishi, T. Taychatanapat, K. Watanabe, T. Taniguchi, A. Yacoby, and P. Jarillo-Herrero, Phys. Rev. Lett. {\bf 108}, 076601 (2012).
\bibitem{sbg} A.L. Grushina, D.-K. Ki, and A.F. Morpurgo, Appl. Phys. Lett. {\bf 102}, 223102 (2013).
\bibitem{mucha} M. Mucha-Kruczy\'nski, E. McCann, V.I. Fal'ko, Semicond. Sci. Technol. {\bf 25}, 033001 (2010).
\bibitem{jacak} L. Jacak, P. Hawrylak and A. W\'ojs, Quantum Dots, Springer-Verlag (1998).
\bibitem{brey} L. Brey and H. A. Fertig, Phys. Rev. B {\bf 73}, 235411 (2006).
\bibitem{koskinen} P. Koskinen, S. Malola, and H. H\"akkinen, Phys. Rev. Lett. {\bf 101}, 115502 (2008).
\bibitem{ost} J.A.M. van Ostaay, A.R. Akhmerov, C.W.J. Beenakker, and M. Wimmer, Phys. Rev. B {\bf 84}, 195434 (2011).
\end{thebibliography}
\end{document}